\documentclass[prb,twocolumn]{revtex4-1}
\usepackage{times}
\usepackage[dvips]{graphicx}
\usepackage{latexsym,amsmath,amssymb,bm,euscript}
\usepackage[dvips]{color}
\usepackage{multirow}

\begin{document}

\title{Monte Carlo study of a $U(1)\times U(1)$ system with $\pi$-statistical interaction}
\date{\today}

\author{Scott D. Geraedts}
\author{Olexei I. Motrunich}
\affiliation{Department of Physics, California Institute of Technology, Pasadena, California 91125, USA}

\begin{abstract}
We study a $U(1)\times U(1)$ system with two species of loops with mutual $\pi$-statistics in (2+1) dimensions. We are able to reformulate the model in a way that can be studied by Monte Carlo and we determine the phase diagram. In addition to a phase with no loops, we find two phases with only one  species of loop proliferated. The model has a self-dual line, a segment of which separates these two phases. Everywhere on the segment, we find the transition to be first-order, signifying that the two loop systems behave as immiscible fluids when they are both trying to condense. Moving further along the self-dual line, we find a phase where both loops proliferate, but they are only of even strength, and therefore avoid the statistical interactions. We study another model which does not have this phase, and also find first-order behavior on the self-dual segment. 
 
\end{abstract}
\maketitle


Systems with statistical interactions between particles arise in many contexts.  For example, Laughlin quasi-particles in the fractional Quantum Hall Effect have mutual statistics which depends on the particular state.\cite{Stern2008}  As another example, spinon and vison excitations in $Z_2$ fractionalized phases in quantum magnets have mutual $\pi$ statistics. 
\cite{Read1989, Kitaev2003, SenthilFisher_Z2}

While gapped topological phases are well understood, the phase transitions between them are less explored.  One reason is that the character of the transitions is a dynamical question, while in cases with statistical interactions, the system path integral contains generally complex phases and hence has sign problem in direct Monte Carlo (MC) approaches.  

References~\onlinecite{Vidal2009,Tupitsyn2010} considered the toric code model, which can be also viewed as an Ising gauge theory with Ising matter fields. \cite{Fradkin1979} In a formulation containing Ising matter and $Z_2$ fluxes (visons) as elementary particles, the two have $\pi$-statistical interaction, and one cannot simulate large systems in these degrees of freedom.  On the other hand, the matter-gauge field formulation does not have the sign problem and was studied in detail in MC.\cite{Tupitsyn2010}
 This system has two phases:\cite{Fradkin1979} deconfined and confined.  The confined phase includes both the Higgs regime (crudely viewed as condensation of matter out of the deconfined phase) and confining regime (viewed as condensation of visons), and there is a path connecting the two regimes that does not cross any phase transitions.  Resembling somewhat liquid-gas system, there is also a finite segment of first-order transitions separating the two regimes along the so-called self-dual line, where the $Z_2$ charges and $Z_2$ vortices have identical interactions.  We can think of the two species as being immiscible on the first-order segment.  Reference~\onlinecite{Tupitsyn2010}  focused on the transition near the tip of the deconfined phase along the self-dual line and possible scenarios how the Higgs and confinement transition lines join.

In this paper, we study a $U(1) \times U(1)$ system with $\pi$ statistical interactions, which appears in effective field theories of frustrated quantum antiferromagnets\cite{Senthil_theta, Xu2009, Kou2008} and other areas.\cite{Cho2011, Hansson2004} We consider concrete lattice realizations that can be reformulated in a sign-free manner and explore these using MC.  

Specifically, we consider a system of two loops with short-range interactions and $\theta = \pi$ statistics,
\begin{equation}
S = \sum_r \frac{\vec{J}_1(r)^2}{2 t_1}
+ \sum_R \frac{\vec{J}_2(R)^2}{2 t_2}
+ i \theta \sum_r \vec{J}_1(r) \cdot \vec{a}_2(r) ~.
\label{action}
\end{equation}
Here $\vec{J}_1$ and $\vec{J}_2$ represent conserved integer-valued currents, $\vec{\nabla} \cdot \vec{J}_1 = \vec{\nabla} \cdot \vec{J}_2 = 0$, residing on the links of inter-penetrating cubic lattices \cite{Fradkin_SL2Z}. The last term is a statistical interaction where the cross-linking of the two loop systems is calculated with the help of an auxiliary ``gauge field'' $\vec{a}_2$ whose flux encodes the $\vec{J}_2$ currents, $\vec{J}_2 = \vec{\nabla} \times \vec{a}_2$.  Just like in the $Z_2 \times Z_2$ case, we can reformulate the model as a special matter-gauge system amenable to MC study.

\begin{figure}
\includegraphics[angle=-90,width=\linewidth]{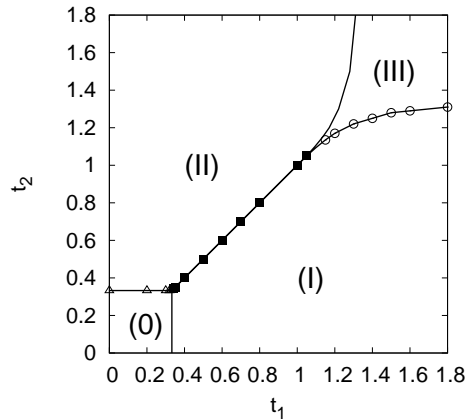}
\vspace{-20pt}
\caption{The phase diagram for the unrestricted model. Phase (0) contains no loops. Phase (I) contains proliferated loops in $J_1$ and no loops in $J_2$, while in phase (II) the variables are interchanged. Phase (III) contains proliferated double-strength loops in both variables. }
\label{phase1}
\end{figure}

Figure~\ref{phase1} is our main result and shows the phase diagram of this model.  For small $t_1$ and $t_2$, there are only small loops (phase 0 in the figure).   If we fix $t_2$ and increase $t_1$, the $J_2$ loops remain small, while beyond some critical $t_1$ coupling, $J_1$ loops condense via an XY transition (phase I); we get another phase (II) if we keep $t_1$ small and increase $t_2$.  Unlike the Ising case, the two phases I and II are distinct, and the following question arises.  Suppose we are increasing $t_1$ and $t_2$ along a self-dual line $t_1 = t_2 = t$ where the two species have identical interactions, so both equally want to proliferate at some point.  One possibility is that they behave as immiscible fluids and phase separate.  Another possibility is a regime where both loops are present in some critical soup, which would be an example of an unusual phase transition.
Such a question is of much recent interest. 
\cite{deccp_science, deccp_prb, SU2ring1, *Sandvik2010, SU2ring2, Alet_dimers3D, Chen2009, Charrier2010, Senthil_theta, Kuklov06, Smiseth2005,Kragset06,Herland2010,Lou2009,Banerjee2010,*Banerjee2011,KamalMurthy,shortlight,Kuklov08,Motrunich,Kaul2011}
 
In the present study we find that in the above specific model, the first scenario happens and we have first-order transition between phases I and II.  Interestingly, if we continue increasing $t_1$ and $t_2$, the two loops eventually condense simultaneously but only in even strength for both $J_1$ and $J_2$, while there are no large loops of odd strength for either species.  By going into this ``paired-$J_1$'' and ``paired-$J_2$'' phase (labelled III in Fig.~\ref{phase1}), the loops avoid the destructive interference effects of the statistical interaction.

Returning to the transition line I-II, the first-order transition is strongest near the two ends of the segment where respectively phase I or III opens up.  We also explore what happens when we modify the model to eliminate phase III, which we realize by simply restricting $|J_1|\leq 1$, $|J_2| \leq 1$, thus preventing pairing within each species.  
The strength of the first-order character indeed decreases as we increase $t$, but for all such accessible values the transition remains first order.

{\it Model and Monte Carlo Method.} In order to reformulate Eq.~(\ref{action}) in a sign-free way, we pass from $J_1$ variables to conjugate 2$\pi$-periodic phase variables by formally writing the constraint at each $r$: 
\begin{equation}
 \delta[\vec{\nabla} \cdot \vec{J_1}(r)=0]=\int_{-\pi}^{\pi} d\phi_{r} \exp[-i \phi_{r} (\vec{\nabla} \cdot \vec{J_{1}})]. 
\label{constraint1}
\end{equation} 
To be precise in our system with periodic boundary conditions, we also require total currents of $J_1$ and $J_2$ to vanish. In this case we can write $\vec{J}_2=\vec{\nabla} \times \vec{a}_2$ and the action~(\ref{action}) is independent of the gauge choice for $a_2$. We enforce the zero current in $J_1$ with the help of fluctuating boundary conditions for the $\phi$-s across a single cut for each direction $\mu=x,y,z$
\begin{equation}
\delta(\sum_{r}J_{1\mu}(r)\delta_{r_{\mu},0})=\int_{-\pi}^{\pi} d\gamma_{\mu} \exp[-i \gamma_{\mu} \sum_r J_{1\mu}(r) \delta_{r_{\mu},0}] .
\label{constraint2}
\end{equation} 

This gives the following partition function:
\begin{equation}
Z= \sum_{{\rm constrained~}\vec{J_2}} \int_{-\pi}^{\pi} \prod_r d\phi_{r} \int_{-\pi}^{\pi} \prod_{\mu=1}^{3}  d\gamma_{\mu} 
e^{-S[\phi,\gamma,a_2]}
\label{Z2}
\end{equation}
where the action is given by:
\begin{eqnarray}
S[\phi,\gamma,a_2]&&=\sum_r \frac{[\vec{\nabla} \times \vec{a}_2(r)]^2}{2t_2}
\label{action2}\\
&&+\sum_{r,\mu} V_{{\rm Villain}}[\phi_{r+\mu}-\phi_r-\theta a_{2\mu}(r)- \gamma_{\mu} \delta_{r_{\mu},0}]
\nonumber
\end{eqnarray}
$V_{{\rm Villain}}$ is the `Villain potential', which is obtained by summing over the $J_1$ variables:
\begin{equation}
\exp[-V_{{\rm Villain}}(\alpha,t_1)]=\sum_{J_{1}=-\infty}^{\infty} \exp \left[-\frac{J_{1}^2}{2t_1}+iJ_1\alpha \right]
\label{Villain}
\end{equation}

In the actual Monte Carlo, we use $\phi_r$, $\gamma_{\mu} \epsilon (-\pi,\pi)$, $a_{2\mu}(r) \epsilon \mathbb{Z}$, and perform unrestricted Metropolis updates. One can show that physical properties measured in such a simulation are precisely as in the above finitely defined model. 

In this work, we monitor ``internal energy per site'', $\epsilon=S/L^3$, and compute heat capacity, defined as 
\begin{equation}
C=(\langle \epsilon^2 \rangle - \langle \epsilon \rangle ^2) \times {\rm Vol},
\label{C}
\end{equation}
where Vol=$L^3$ is the volume of the system. 
To determine the phase diagram, we monitor loop behavior by studying ``superfluid stiffness'', which is defined for loops of flavor $a$ as:
\begin{equation}
\rho^{\mu\mu}_a(q)=\frac{1}{{\rm Vol}}\left \langle \left| \sum_{r} J_{a\mu}(r) e^{i\vec{q} \cdot \vec{r}} \right|^2\right \rangle.
\label{rhoE}
\end{equation}
Because of the vanishing total current, we define these at the smallest non-zero $q$; e.g., for $\rho^{xx}$ we used $\vec{q}=(0,\frac{2\pi}{L},0)$ and $\vec{q}=(0,0,\frac{2\pi}{L})$. We focus on flavor 2 since it is more readily accessible in the formulation of (\ref{action2}). 
We also monitor gauge-invariant ``magnetization'', defined as:
$M=\sum_r e^{2i\phi_r}$,
which can detect flavor 1 condensates (care is needed interpreting $M$ for different boundary conditions).

\begin{figure}
\includegraphics[angle=-90,width=0.95 \linewidth]{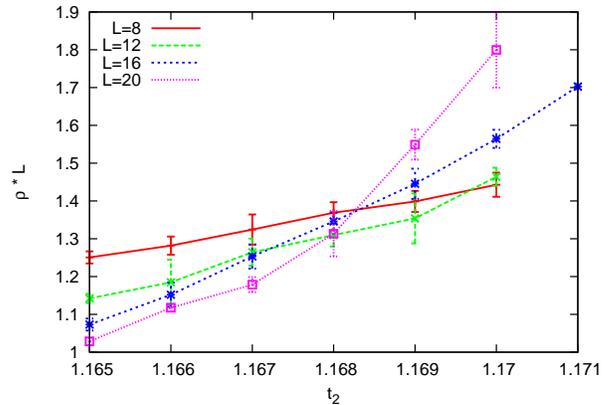}
\vspace{-10pt}
\caption{$\rho_2(q_{min}) \cdot L$ as a function of $t_2$ at $t_1=1.2$. Based on this data, we conclude that the phase transition occurs at approximately $t_2=1.168$, and is second order in nature. Scans like this were used to determine the phase diagram in Fig. \ref{phase1}. Here the transition is between (I) and (III). Each data point is the result of $5 \times 10^6$ Monte Carlo sweeps. Error bars were determined by looking at the difference between runs with different initial conditions.}
\label{rho}
\end{figure}

{\it Results for Unrestricted Currents.} We determined the phases of this model by looking at the order parameters $\rho_2$ and $M$. $\rho_2$ is non-zero in phases (II) and (III), and $M$ is non-zero in phases (I) and (III). We found the locations of the phase boundaries more accurately by studying $\rho_2 \cdot L$ crossings. We took data in sweeps across the phase diagram (see Fig.~\ref{phase1}), and defined the intersection of the $\rho_2 \cdot L$ curves to be the location of the phase transition. An example of such a sweep is shown in Fig.~\ref{rho}. The sweep is `vertical' in the phase diagram, i.e.\  fixed $t_1$. The symbols on the phase diagram are the values of $t_1$ at which the sweeps were performed.
We studied the fine nature of the phases by looking at clusters formed by $J_2$ loops. In phase (II), the largest clusters of $J_2$ grow with system size, and have arbitrary $J_2$, so we deduced that this phase contains condensed loops of $J_2$. In phase (III), the loops that scale with system size have even $J_2$, so we judged that the condensed loops have even strength. In phases (0) and (I), we found no large clusters, and deduced that $J_2$ is gapped. The model with $t_1=0$ is a model containing only one species of loop\cite{Cha}. Our value for the position of the (0)-(II) XY transition ($t_2 \approx 0.333...$) is in agreement with prior work on this model\cite{Sorensen}. For $t_1 \rightarrow \infty$, the Villain weight (\ref{Villain}) vanishes except for $\alpha=2\pi \times$ (int), which enforces $J_2=\vec{\nabla} \times \vec{a}_2=2\times$(int). Therefore, at large $t_1$ the (I)-(III) transition is a transition from no loops of $J_2$ to loops of even $J_2$. One expects that this transition is XY-like, and similar to the (0)-(II) transition, but due to doubled $J_2$, it should occur at a $t_2$ value four times higher. We observed the (I)-(III) transition to occur at $t_2 \approx 1.3$ for large $t_1$, in agreement with this expectation. 

\begin{figure}
\includegraphics[angle=-90,width=0.95 \linewidth]{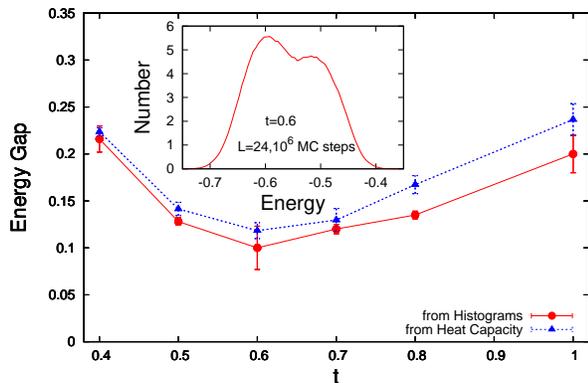}
\vspace{-20pt}
\caption{Plot of the energy difference between the peaks of the histograms, as a function of $t_1=t_2=t$ on the self-dual line. Higher values denote a stronger first-order nature. The inset shows the energy histogram for $t_1=t_2=0.6$. The dual-peaked shape of the histogram implies a first-order transition. The histograms were performed at $L=24$ with $10^6$ Monte Carlo sweeps. The heat capacity intercepts were found using data from $L=8,10,12,14$ and $16$.}
\label{order1}
\end{figure}

To determine whether the loops are immiscible, or if a critical state is possible, we consider whether the phase transition (I)-(II) on the self-dual line is first- or second-order. We studied this by looking at histograms of $\epsilon$. In the continuous case, these histograms would be singly-peaked, while in the first-order case we expect to see two distinct peaks. An example of such a histogram is shown in the inset for Fig.~\ref{order1}. We observed dual-peaked histograms for all points on the self-dual line and concluded that the transition is first-order. One way to quantify the strength of the transition is to look at the distance between the two peaks in these histograms. We plotted this peak-to-peak distance in Fig.~\ref{order1}, which is clearly non-zero for all $t$. 

We can also determine the order of the transition by looking at the heat capacity, where the signature of first-order is 
$C$ growing as $L^3$. Phenomenologically\cite{Chala1986}, at a first-order transition the energy distribution is described by a sum of two normal distributions, $P(\epsilon)=c_+p_++c_-p_-$, where $c_+$ and $c_-$ are the weights representing how much time the system spends in each state, and $p_+$ and $p_-$ are two normal distributions centered at $\epsilon_+$ and $\epsilon_-$. Computing heat capacity based on this ansatz gives:
\begin{equation}
\frac{C}{{\rm Vol}}=c_+c_-(\epsilon_+-\epsilon_-)^2+\frac{B}{{\rm Vol}}=A+\frac{B}{{\rm Vol}},
\end{equation}
where $\epsilon_+-\epsilon_-$ is the peak-to-peak distance, and $B$ is a volume-independent constant. 
We plotted $\frac{C}{{\rm Vol}}$ vs.\ $\frac{1}{{\rm Vol}}$ and found the y-intercepts of these plots, which should be equal to $A$. If we assume that $c_+=c_-=0.5$ as a rough estimate, then the energy gap should be equal to $2\sqrt{A}$. The resulting estimate from heat capacity is plotted in Fig.~\ref{order1}. Both of our measurements of the energy gap show that the transition is strongly first-order at $t=0.4$. The strength of the transition initially decreases with increasing $t$, before increasing again after $t \approx 0.6$. Thus, in this model the two condensates are immiscible. 

In addition to finding the position of the phase boundaries, plots of $\rho_2 \cdot L$ like the one in Fig.~\ref{rho} can be used to study the nature of the (I)-(III) phase transition. We can argue that at high $t_1$, the phase transition is continuous. We are interested in whether the nature of the transition changes to first order before it meets up with the transition on the self-dual line. In a first-order transition, we expect the values of the $\rho_2 \cdot L$ crossings to increase with $L$. We do not observe this in Fig.~\ref{rho}, or plots at other $t_1$. We therefore suspect that the transition is second-order, though our data is not precise enough to rule out a weak first-order transition. We also obtained histograms of the energy and magnetization at the phase transition located by the $\rho_2 \cdot L$ plots. We found no evidence of two peaks at $L=16,20$. Since we do not know the critical point exactly, it is difficult to study such histograms at larger sizes, so once again the data indicates a second-order transition but cannot rule out a weak first-order one. 

{\it Results for the Model with Restricted Currents.} From Fig.~\ref{order1}, we have seen that the first-order character of the phase transition on the self-dual line is strongest where it meets phases (0) and (III). Therefore we might expect the transition to be more weakly first-order if we could eliminate phase (III). This is the reasoning behind the restricted model, where we only allow loops with $|J_1| \leq 1$, $|J_2| \leq 1$. This has been done in the Monte Carlo by only changing $a_2$ if the resulting curl satisfies the restrictions, and by restricting the sum in Eq.~(\ref{Villain}) to run only over the values $-1,0,1$. 

\begin{figure}
\includegraphics[angle=-90,width=0.95 \linewidth]{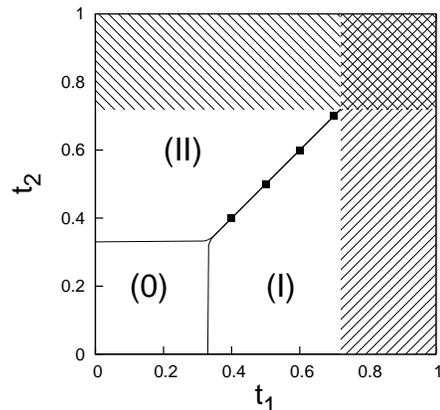}
\vspace{-15pt}
\caption{The phase diagram of the `restricted' model. The region with +45$^{\circ}$ lines is inaccessible in the formulation~(\ref{action2}), and the cross-hatched region is inaccessible in this work. The transition is first order everywhere on the accessible portion on the self-dual line, but the strength decreases as $t$ is increased.}
\label{phase2}
\end{figure}

Figure~\ref{phase2} shows the phase diagram for this modified system, which was determined using the same methods as Fig.~\ref{phase1}. The boundaries of phase (0) are very similar to the unrestricted model, which is not surprising as at the (0)-(I) transition the proliferating loops are mostly of strength 1. Referring to the Villain potential~(\ref{Villain}), we see that if the sum on the right side is negative, the potential is undefined. When the sum is restricted to $|J_1| \leq 1$, this occurs for $t_1 \gtrsim 0.72135$, and this limits the area of the phase diagram that we can study with this formulation. Upon using interchange symmetry, Fig.~\ref{phase2} contains a region inaccessible in this work, indicated by cross-hatching. 

\begin{figure}
\includegraphics[angle=-90,width=0.95 \linewidth]{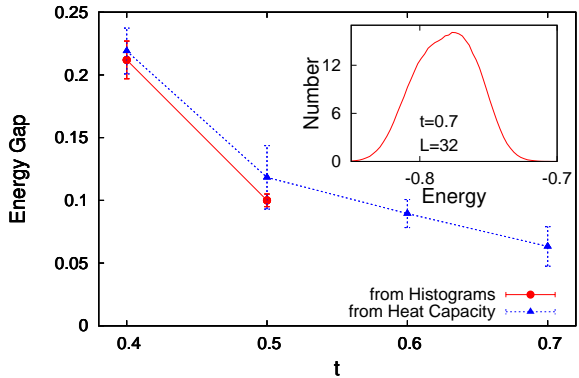}
\vspace{-20pt}
\caption{Same as figure \ref{order1}, but for the restricted model.}
\label{order2}
\end{figure}

We now investigate whether removing phase (III) has changed the nature of the transition on the self-dual line. We used energy histograms and studies of the heat capacity to determine the peak-to-peak distance. The heat capacity suggests a first-order transition. For $t=0.6, 0.7$, the energy gaps were too small to be accurately determined by studying the histograms. The histogram for $t=0.7$ is shown in the inset to Fig.~\ref{order2}. We cannot resolve two separate peaks, but the distribution has a flat top, which suggests that the transition is weakly first-order. In order to acquire more clearly two-peaked histograms, we studied the magnetization of the system. We found the peaks in these to be more easily distinguished, and the results clearly indicate a first-order transition. 

{\it Discussion.} We studied a lattice realization of a $U(1) \times U(1)$ system with $\pi$-statistical interactions.  It was helpful to know the location of the phase transition between I and II from self-duality in this non-trivial 3D Statistical Mechanics problem.  In two somewhat different models, we found first-order transitions on the self-dual line, which means that when both loops are trying to condense, they tend to phase-separate. A continuous transition would be an example of an NCCP$^1$ $U(1)$ self-dual critical point \cite{Senthil_theta}. We found in the restricted model that the first-order transition became weaker as $t$ was increased, but we could not study the model for $t$ higher than a certain value. If one could find a way to study the model at high $t$, it would be interesting to see if the first-order transition continues to weaken and perhaps becomes second-order. One could also explore more models asking if some short-range modifications can produce a critical loop state. There is evidence for a continuous transition in $SU(2)$ spin models, \cite{SU2ring2,Sandvik2010,Lou2009,Banerjee2010,*Banerjee2011} but our system has no analog to these. 

Our study is an example of a sign-free reformulation of a model with statistical interactions and the power hence afforded by Monte Carlo to establish the phase diagrams and study phase transitions. Though we determined most of the phase diagram in good detail, it may be useful to get a better understanding of how the phase transitions join at the corners of the (0) and (III) phases.  It would also be interesting to explore more models with statistical interactions that can have such reformulations.  An accessible direction already in the present setting is to examine the model Eq.~(\ref{action2}) with general statistical angle $\theta$.  Another interesting direction is to introduce some attraction between the two loop species, to see if we can achieve fermionic bound states of $J_1$ and $J_2$ and what phases can be accessed in this way.

{\it Acknowledgements. }We are grateful to A.~Vishwanath, M.~P.~A. Fisher, and T.~Senthil for many stimulating discussions, and in particular thank A.~Vishwanath for careful reading of the manuscript and useful suggestions. This research is supported by the National Science Foundation through grant DMR-0907145, and by the XSEDE computational initiative grant TG-DMR110052.

\bibliography{Interacting_Loops}
\end{document}